\documentclass[nofootinbib,twocolumn]{revtex4}

\usepackage[pagewise]{lineno}
\usepackage{graphics}
\usepackage{graphicx}

\begin{document}

\title{Properties of the Schwinger series and pair creation in strong 
fields}

\author{ B.\,B. Levchenko }
\email{levtchen@mail.desy.de}

\affiliation{Skobeltsyn Institute of Nuclear Physics,
M.V. Lomonosov Moscow State University,\\
 119991 Moscow, Russian Federation}


\begin{abstract}
Probabilities of a pair of fermions and  bosons  creation  in a static and spatially  uniform electric 
field $E$ are represented in the Schwinger formulas by infinite series. 
It is believed that in weak fields the main contribution to the probability 
  is given by the first term of series, however the  size of the 
 remainder apparently was analyzed by nobody. 
 We study the mathematical structure of the Schwinger series 
 by using methods developed during last decades and  prove that the given 
series allows an  exact summation and the contribution of 
remainder growths rapidly with the field strength. 
 As a rule, it is argued that
the pair of particles begin to be produced efficiently 
from the vacuum only in the fields of strength  $E\geq E_{cr}$.
 However, the direct calculation shows  that the Schwinger formula for 
 creation of $e^+e^-$ pairs  is valid only  at the field intensities 
of $E<0.0291E_{cr}$.  At  higher fields, the probability of pair production 
in an unit space-time volume
exceeds unity.  In this regard, we refine 
the formula for the probability of pair creation 
and numerically  find that in the field of strength $2.95\%$ of $E_{cr}$ the  pair
production probability  is almost 100$\%$.
\end{abstract} 


\maketitle


\textbf{1.} 
More than sixty years ago Julian Schwinger~\cite{schw51} published the formulas 
that allow to calculate the effects from impact of strong electromagnetic fields on the  vacuum. 
Were studied the vacuum polarization in strong electric~${\bf E}$ and magnetic ~${\bf H}$ 
fields, and the resulting one-loop nonlinear correction to the Maxwell's Lagrange function.
Of this fundamental paper on quantum electrodynamics we are interested in a
 particular result  obtained  "incidentally"  by a modest statement of  Schwinger~\cite{schw51}.
It was found that the change in the Lagrangian density causes the appearance of
imaginary part of the effective action that corresponds to a non-zero probability
of actual pair creation in the field.
The probability $\bar{w}^{(s)}$, per unit time and per unit volume, 
that a pair of particles carrying spin $s$ is created
by a  homogeneous and static electric field is expressed by~\cite{schw51} 
\begin{equation}
\bar{w}^{(s)}(E)=(2s+1)\frac{\alpha}{2\pi^2}   E^2 \sum_{n=1}^{\infty} 
\frac{(\pm 1)^{n+1}}{n^2} e^ { -\pi n E_{cr}/E  }\, ,
\label{1.1}
\end{equation}
where\footnote{Schwinger employs units in which $\hbar = c = 1$.}   
   $\alpha= e^2/4\pi$, $E_{cr}={m}^2/e$ is the so-called critical field,
$m$ the particle mass, $e$ the charge on the particle.
The '+' sign corresponds to  creation of fermion pairs, and the '-' sign for  production 
of bosons. At production of  $e^+e^-$ pairs,  $E_{cr}=1.32\times 10^{16}$\,V/cm.
The scale factor  $E_{cr}$ was introduced and evaluated for the first time in 1931 by Fritz Sauter \cite{sau31}. 
Equation  (\ref{1.1}) corresponds to the  weak field approximation.

The peak 
electric field achieved at the focus of today's most powerful lasers 
still several orders of magnitude lower 
$E_{cr}$.
This is the main obstacle for a direct experimental verification of given
 fundamental mechanism of particle production.
Despite of this, the  Schwinger mechanism from QED, was the basis for successful phenomenological
models of hadron production in collisions at high energies via the tunnel creation
 of $ q \bar{q} $-pairs in string-like  chromoelectric field \cite{casher,lund1,lund2}.

From the form in which presented the result (\ref{1.1}) and the value of $ E_{cr} $, it seems
self-evident that in weak fields the main contribution to the production probability 
gives the first term of the series
\begin{equation}
\bar{w}^{(1/2)}\approx \frac{\alpha}{\pi^2} E^2\exp \Big \{ -\pi\frac{E_{cr}}{E} \Big \},
\label{1.2}
\end{equation}
and this probability is very small. It should be noted, however, that the total value of
residual terms of an infinite series, apparently not analyzed.
One of the purposes of this article\,\footnote{ Based on a report presented 
at the conference "Lomonosov Readings", MSU, Moscow, Russia, 14 Nov  2011.}
is an evaluation how strongly change the  remainder of the series (\ref{1.1})
with increasing field strength .
The analysis of this formal mathematical problem allowed to find the exact sum
of the infinite series (\ref{1.1}), and find a number of unexpected consequences,
not mentioned by other authors.


\vspace*{2mm}
\textbf{2.} To solve the above problem we introduce the reduced field
$ \beta = E / E_ {cr} $, and denote the exponential factor in (\ref{1.1})
 by $x=\exp (-\pi/\beta )$. As a result,  (\ref{1.1}) takes the form
\begin{equation}
\bar{w}^{(s)}= \pm (2s+1)\frac{\alpha}{2\pi^2} E^2_{cr} {\beta}^2 
\sum_{n=1}^{\infty} \frac{(\pm x)^n}{n^2}.
\label{dl.0}
\end{equation}
Analysis of the mathematical literature has shown that the series in (\ref{dl.0})
belongs to a class of polylogarithms, and in our particular case we are dealing with the  dilogarithm.
In mathematics, the dilogarithm studied since Leibniz and Euler, but in the last quarter
of the 20th century polylogarithms and hyperlogarithm again attracted attention and
appeared in many branches of mathematics and physics. Polylogarithms and  reference  information
of them are included in the standard set of special functions in most packages
of applied mathematical programs.

\begin{table}[t]
\caption{The exact values of the Rogers dilogarithm }
\centering
\begin{tabular}{ccccccccc}
 \hline
 \hline
&&&&&&&&\\
$x$ & $-\infty$ & $-1$ & $0$ &$\frac{(\sqrt{5}-1)^2}{4}$ & {\large $\frac{1}{2}$} & $\frac{\sqrt{5}-1}{2}$& $1$ & $\infty$   \\
&&&&&&&&\\
&&&&&&&&\\
$\mathrm{L}(x)$ & {\large $-\frac{\pi^2}{6}$ \ } &{\large $-\frac{\pi^2}{12}$}\  
&\ $0$\ &\ {\large $\frac{\pi^2}{15}$ }\  &\ {\large $\frac{\pi^2}{12}$ }\ &\ {\large $\frac{\pi^2}{10}$ }\  &\ {\large $\frac{\pi^2}{6}$}\ 
&\ {\large $\frac{\pi^2}{3}$ }\ \\ 
&&&&&&&&\\
\hline
\hline
\end{tabular}
\label{table_dlv}
\end{table}

Let us write out the definition of the dilogarithm and its integral representation 
following the interesting original review \cite{kiril94},
\begin{equation}
\mathrm{Li_2} (x)=\sum_{n=1}^{\infty} \frac{x^n}{n^2}=
-\int_0^x\frac{\ln(1-t)}{t}\mathrm{d} t .
\end{equation}
Thus, in terms of the function $ \mathrm {Li_2} (x) $ the probability 
of pair creation we rewrite as follows,
\begin{equation}
\bar{w}^{(s)}(\beta)= \pm (2s+1)\frac{\alpha}{2\pi^2} E^2_{cr} {\beta}^2 
\mathrm{Li_2}(\pm e^{-\pi/\beta }) ,
\label{dl.1}
\end{equation}
and the problem of summing the infinite series in (\ref{1.1}) can be regarded as solved.
However, to clarify the meaning of this result, let us recall some properties of 
the dilogarithm \cite{kiril94}.
In this case it is more convenient and more compact to use the Rogers dilogarithm
$ \mathrm{L}(x)$, $0<x<1$,  defined as
\begin{equation}
\mathrm{L}(x)=\mathrm{Li_2}(x) + \frac{1}{2}\ln (x)\ln (1-x) .
\end{equation}
The function $ \mathrm {L} (x) $ 
allows an analytical continuation into the complex plane with 
cuts $ (- \infty, 0] $ and $ [1, + \infty) $ along the real axis.
Using a number of theorems and functional relations, $ \mathrm {L} (x) $ can be 
extended to the whole real axis. 
There are only few real points $ \{x_i \} $, where the values of $ \mathrm {L} (x) $ 
are known exactly. Their values are summarized in Table I.
In general, for more than two centuries of research were opened a large number of functional identities, 
allowing to relate dilogarithms  with  different powers of $x$. 
There are identities and for more complicated rational arguments.

Coming back to the problem of calculating the probability of pair creation by  (\ref{dl.1}), 
we summarize that by using the integral representation of $ \mathrm{Li_2}(x) $, 
one can calculate $ \bar{w}^{(s)} (\beta) $ with any given accuracy, controlled by the precise values 
of $ \mathrm{L} (x_i) $ from Table I. Another approach is possible also.
 Having determined numerically $ \mathrm{L} (x) $ for a given $ x_0 $, 
 the values of $ \mathrm {L} (x) $ at other points are restored through a chain of functional identities.

To conclude this section we estimate, at which values of $ \beta $ the 
remainder of the series (\ref{1.1}) starts to give a significant 
increment to (\ref{1.2}). We consider only the case of $ e^+ e^- $ pairs. 
To do this, compose a ratio of the total Schwinger series to its first term
\begin{equation}
R(\beta)= \frac{\mathrm{Li_2}( e^{-\pi/\beta})}{e^{-\pi/\beta}}.
\end{equation}
By a numerical integration one find (see Fig. 1) that only at $\beta> $ 0.5 
the remainder of the series begins to play an important role, making 
the correction of $6\%$ at $ \beta = 2 $. 
But these values of $ \beta $ are beyond the weak field approximation. 
Thus,  (\ref{1.2})  indeed is a very good approximation to (\ref{1.1}).


\begin{figure}[t]
\begin{center}
\includegraphics*[width=1.0\columnwidth]{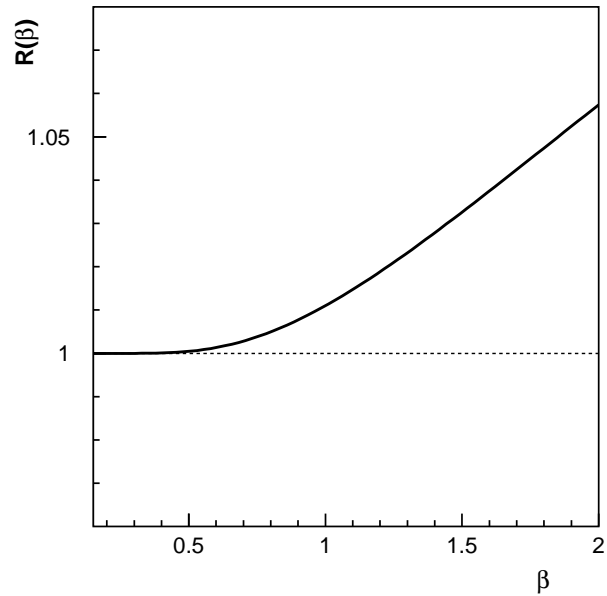}
\caption{$e^+e^-$ pair production.
 The ratio of the total Schwinger series (\ref{1.1}) to its first term (\ref{1.2}) as a 
function of the reduced field $ \beta $.}
\label{Fig:1}
\end{center}
\end{figure}


\vspace*{2mm}
\textbf{3.} 
The method of the simplified record of results 
when a group of some physical constants are set equal to unity
is widespread in theoretical physics.
In the case under study (\ref{1.1}) it is $ \hbar = c = 1 $.
As noted by several authors \cite{vlas90}, "simplifications"
achieved in the natural system of units (the system of Planck and Hartree, a  relativistic system), 
sometimes are fraught with illusions\,\footnote{ 
In this regard, in the book \cite{vlas90} on page 97 there is a remarkable comment:
"Thus, simplification of the formulation of physical laws by using the natural 
systems of units is bought at the cost of replacing the physical equations 
by numerical one.  In the used units,  the numerical equations relates among 
themselves  not the physical quantities but only their numerical values. 
 A possibility of comparison and check of the dimension of the considered 
 physical quantities is lost. Due to preservation of the same lettering 
a substitution of physical equations by the numerical one 
 is masked and there is an illusion of operating with real physical 
quantities and equations. 
 Arises abstractness of used equations and the theory  as a whole."} 
and  Schwinger formula (\ref{1.2})  may serve as a vivid illustration.

In the vast literature devoted to  pair production  in electromagnetic fields, 
we have not ever met a graphical representation of  (\ref{1.2}).
Before we draw the graph $ \bar{w}^{(1/2)} $ as a function of $ \beta $, 
one need to choose the system of units.
In  \cite{schw51} Schwinger 
 likely 
used the Heaviside system of units, the most common 
in scientific literature of the time. In (\ref{1.2}) the dimension of the probability 
density is cm$^{-3}\cdot$s$^{-1}$ and the dimension of electric field V/cm. 
We need to restore the "invisible" conversion factor  to match the dimensions on both sides.
Let us reproduce schematically the logic of calculations led to
(\ref{1.1}) in order to recover all constants $ \hbar $ and $c$
and bring the final result in the SI units\,\footnote{However, in the following, 
as the unit of length is more convenient to use 1\,cm.}.

Suppose we observe  during a period of time $ \Delta T$ for a three dimensional space 
volume $ \Delta V$  
filled with a homogeneous and static electric field $E$.
We write the wave function of the vacuum as $\Psi=\exp(iS/\hbar)$,
where the action $S$ is related to the density of the Lagrangian function
 $\cal{L}$ via $S={\cal{L}}\,\Delta V\Delta T$.
 As a result, the probability that the vacuum will be stable is 
$W_v=|\Psi|^2 = \exp(-2\mathtt{Im}\,{\cal{L}}\,\Delta V\Delta T/\hbar)$,
and the probability of pair creation in the field is
\begin{equation}
W_p=1 - W_v = 1- \exp(-\frac{2}{\hbar}\mathrm{Im}\,{\cal{L}}\,\Delta V\Delta T).
\label{3.1}
\end{equation}
In the weak-field approximation, keeping only the first term in the expansion 
of the exponential, we reproduce the  Schwinger result, 
\begin{equation}
W_p^{(s)} \approx \frac{2}{\hbar}\mathrm{Im}\,{\cal{L}}\,\Delta T\Delta V
\,,
\label{3.1.1}
\end{equation}
 that is, the probability of pair production per unit time per unit volume is
$\bar{w}^{(s)}=2\mathtt{Im}\,{\cal{L}}/\hbar$.
The dimensions of the functions  $S$ and $\cal{L}$ are fixed by the above 
formulas allowing to restore the required number of the desired constants.
Thus, 
$E_{cr}=m^2c^3/e\hbar$ and 
\begin{equation}
 \bar{w}^{(1/2)}(\beta)= w_0 {\beta}^2  e^{-\pi/\beta}
\label{3.2}
\end{equation}
where
$$ w_0= \frac{e^2}{4\pi\hbar c}\frac{E^2_{cr}}{\pi^2 \hbar}=\frac{m^4c^5}{4{\pi}^3{\hbar}^4}\,.$$
For $e^+e^-$ pairs, after substituting  numerical values of the constants 
one find that in (\ref{1.2})  behind the seemingly small pre-exponential factor 
is "hidden" the very large dimensional  constant 
$ w_0 = 1.087036 \times 10^{50} $ \,cm$^{-3}\cdot$s$^{-1}$. 
Consequently, the earlier qualitative conclusion about the smallness of $ \bar{w}^{(1/2)} $ 
may not be entirely correct\,\footnote{Our expression for $w_0$  coincides with the analogous 
scaling constant of Refs \cite{heyl97, ringw01}.}.

In Fig. 2  by open dots shown $W_p^{(1/2)}$ calculated with  (\ref{3.1.1})  
and  (\ref{3.2}) with $\Delta T\Delta V=1$\,cm$^3\cdot$s.
Taking into account  the magnitude of $w_0$, the result is not so surprising, 
despite the fact that the field  strength is much lower than the critical value.
As follows from Fig. 2,  
the Schwinger formula  is valid only for fields of $ E <\beta_c E_{cr}$.
At higher field, the probability of pair production 
exceeds unity.

From (\ref{3.2}) we find $\beta_c$ at which $ W_p^{(1/2)} = 1 $,
\begin{equation}
 \exp\{\ln ( {\beta_c}^2w_0\Delta T\Delta V) - \frac{\pi}{\beta_c}\} = 1\,.
\end{equation}
Let us denote $A=\frac{\pi}{2}\sqrt{w_0\Delta T\Delta V}$  and sequentially making change of variables 
$ \sqrt{w_0\Delta T\Delta V} \beta_c =z$ and $ z = e^W $, we get an equation
\begin{equation}
W(A)e^{W(A)}=A
\label{lamb}
\end{equation}
 for the Lambert $W(A)$ function.
The Lambert function has the same fate as that of the dilogarithm.
Interest in it was revived in the last quarter of the 20th century
\cite{knuth96}, \cite{dub06}.
The required $ \beta_c$ is obtained from 
$$\beta_c=\frac{1}{\sqrt{w_0\Delta T\Delta V}}e^{W(A)}.$$
For numerical estimates we use the asymptotic expansion of $ W (A) $ \cite{knuth96}
\begin{eqnarray}
W(A)&=&L_1 -L_2+\frac{L_2}{L_1}+\frac{L_2(L_2-2)}{2L_1^2}\nonumber \\
&+&\frac{L_2(2{L_2}^2-9L_2+6)}{6{L_1}^3}+{\cal O}\Big (\Big\{\frac{L_2}{L_1} \Big\}^4 \Big ),
\end{eqnarray}
where $L_1=\ln A$, $L_2=\ln\ln A$. Thus, $\beta_c(W_p=1)  \approx 0.02905$
at $\Delta V\Delta T=1$~cm$^{3}\cdot$s.
%


\begin{figure}[t]
\includegraphics*[width=1.0\columnwidth]{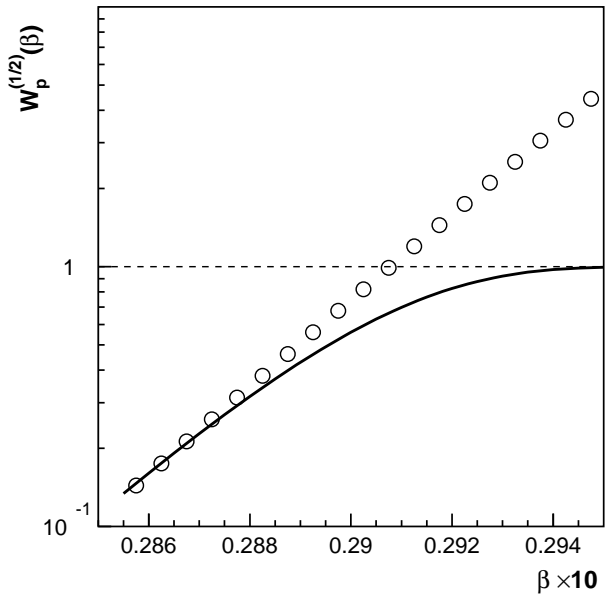}
\caption{The probability of an $e^+e^-$
 pair production in an unit 4-volume, 
$\Delta V\Delta T=1$~cm$^{3}\cdot$s,
 as a function of the reduced field $\beta $.
The open dots shows the result of calculation by equation (\ref{3.1.1}), 
violating the unitarity.
 The solid curve shows the calculation by (\ref{3.tot}). }
\label{Fig.2}
\end{figure}

The reason on which at $ \beta> \beta_c $   the unitarity is violated, lies in the use 
of the weak field approximation leading to (\ref{3.1.1}).
To restore the unitarity, it is necessary to take a step back to (\ref{3.1}), 
which is free of the problem.
In (\ref{3.1}), $2\,\mathrm{Im}\,{\cal{L}}/\hbar$ should be replaced by 
$\bar{w}^{(s)}$ from (\ref{dl.1}) 
and take into account the new system of units (\ref{3.2}).
After completing the necessary substitutions, we obtain the corrected
formula for  the probability of pair creation in the volume  $ \Delta V $
\begin{equation}
W_p^{(s)}(\beta) = 1- \exp\{ \pm (s+\frac{1}{2}) w_0{\beta}^2 
\mathrm{Li_2}(\pm e^{-\pi/\beta })\,\Delta V\Delta T\}
\label{3.tot}
\end{equation}
during the observation period $\Delta T$.
 In (\ref{3.tot}) already makes sense to substitute the exact form we found for 
the Schwinger series, because we are not limited by the weakness of the field\,\footnote{
Provided that the  two-loop correction \cite{ritus75} 
to the Heisenberg-Euler Lagrangian density gives  a small contribution.}.
It should be emphasized that (\ref{3.tot}) explicitly demonstrates how the probability 
of the process depends on the size of the space-time volume.
But this dependence is extremely weak if compared with the dependence on $ E $.
For example, if at a given field strength  the probability of pair creation is 
 $ 95\% $, and we wish to reduce the vacuum field   by the factor ten, keeping the same
particle production rate, then it is necessary to increase the field 
space-time volume  in $10^2\cdot\exp(\frac{9\pi}{\beta}) \approx 10^{420}$ times,
 that is, literally to the cosmological size.

In Fig. 2 by the solid curve shown the probability of an
$ e^+e^- $ pair production calculated by means of (\ref{3.tot})
with $\Delta V\Delta T=1~$cm$^{3}\cdot$s.
As is evident from the figure, $W_p^{(1/2)}$ increases rapidly 
within a very narrow range of variation of $E$.
Indeed, when the field strength is $ 2.75\% $ of $ E_{cr} $ 
the pair production probability is less than $ 1\% $, but in 
the field of strength $ 2.95\% $ of $ E_{cr}$, $e^+e^-$ pairs 
are produced with almost $ 100\% $ probability.
This behavior is typical for tunneling processes.


\vspace*{2mm}
\textbf{4.} 
 In the present 
article we prove that the Schwinger series (\ref{1.1}) for the 
probability of a pair of bosons and fermions production in a uniform and static electric 
field allows to be represented in terms of the dilogarithm.
In a weak field, $ \beta <0.5 $, this series with a very high accuracy is described by 
the first term. With increasing $ \beta $ the
remainder of the series begins to play an important role, making the correction of
 $ 6\% $ at $ \beta = 2 $.
The restoration of all constants $\hbar$ and $c$  in the Schwinger formula reveals 
a very large pre-exponential factor 
$ w_0= 1.087036\times 10^{50}$\,cm$^{-3}\cdot$s$^{-1}$.
Its presence determines a sharp increase  of the pair production  probability 
within a very narrow range of field intesities 
and 
in the weak field approximation leads to a violation of unitarity at $E> 0.0291E_{cr}$.
Equation (\ref{3.tot}) corrects this defect, satisfies the unitarity condition and 
includes the space-time volume of the process.
Numerical estimates show that in the field of strength $2.95\%$ of $E_{cr}$ within 
the four-dimensional volume of 1\,cm$^{3}\cdot$sec, 
the probability of an $e^+e^-$ pair production is close to $100\%$.

\vspace*{2mm}
\begin{acknowledgments}
The author wish to thank S.A. Smolyansky for a reading of the manuscript and comments.

\end{acknowledgments}

\end{document}